\documentclass[a4paper,11pt]{article}
\usepackage{graphicx}
\usepackage{psfrag}
\usepackage{amsmath}
\usepackage{fleqn}
\usepackage{dcolumn}
\usepackage{epsfig}
\usepackage{pslatex}

\newcommand{\avg}[1]{\langle #1 \rangle}

\newcommand{\bid}{\text{bid}}
\newcommand{\ask}{\text{ask}}
\newcommand{\tickVol}{\sigma_\text{tick}}
\newcommand{\annVol}{\sigma_\text{ann}}

\begin{document}
\begin{center}
{\LARGE\bf
   How the trading activity scales \\
   with the company sizes in the FTSE 100
}

\vspace*{4ex}
{\Large
Gilles Zumbach
}

\parbox[t]{0.35\textwidth}{
Commerzbank Securities\\
60 Gracechurch Street\\
London EC3V 0HR}\\[2ex]
and\\[2ex]
\parbox[t]{0.4\textwidth}{
Consulting in Financial Research\\
Ch. Charles Baudouin 8\\
1228 Saconnex d'Arve\\
Switzerland \\
e-mail: gilles.zumbach@bluewin.ch
}

\vspace*{3ex}
To appear in: Quantitative Finance, 2004

\vspace*{3ex}
January 17, 2004

\vspace{2ex}
\vspace*{10ex}

{\Large\bf Abstract}
\end{center}
This paper investigates the scaling dependencies between
measures of ``activity'' and of ``size'' for companies
included in the FTSE 100.
The ``size'' of companies is measured by the total market capitalization.
The ``activity'' is measured with several quantities related to trades
(transaction value per trade, transaction value per hour, tick rate),
to the order queue (total number of orders, total value),
and to the price dynamic (spread, volatility).
The outcome is that systematic scaling relations are observed:
1)~the value exchanged by hour and value in the order queue have exponents {\it lower }
than 1 respectively 0.90 and 0.75;
2)~the tick rate and the value per transaction scale with the exponents 0.39 and 0.44;
3)~the annualized volatility is independent of the size,
  and the tick-by-tick volatility decreases with the market capitalization with an exponent -0.23;
4)~the spread increases with the volatility with an exponent 0.94.
A theoretical random walk argument is given that relates
the volatility exponents with the exponents in points 1 and 2.

\vfill
The author would like to thank Xavier Gabaix for many stimulating discussions.

\newpage

\section{Introduction}
\label{sec-intro}
With high-frequency data, the ``activity'' of a given stock can be measured in several ways,
using data either from the flow of orders and transactions,
from the order queue, or from the volatility (measuring price fluctuations).
Using data from the flow of orders and transactions, we measure
the transactions value per hour, the transaction value per transaction,
the number of transaction per tick, and the tick rate.
When using data from order queues, other measures of  ``activities'' are
the number of orders $n_\text{queue}$ and the total order value $v_\text{queue}$ in the queue.
Finally, we measure the ``activity'' according to the price fluctuations
by defining an annualized volatility $\annVol$ and a tick-by-tick volatility $\tickVol$.
For different stocks, with an increasing capitalization, we can expect
that these ``activity'' measures will change in some systematic ways.
The main question we are addressing in this communication is how
these measures of  ``activity'' scale with the company's ``size''.
The natural measure of size is the market capitalization $M$ of each company,
however we also used the number of shares as a measure of the size.

To be more precise, from the flow of orders and transactions,
we define the cumulative number of tick $N$,
the cumulative number of transaction $D$ ('$D$' for deal),
and the cumulated transaction value $V$.
For example, the cumulative transaction value up to time $t$ is defined by
\begin{equation}
    V(t_0, t) = \sum_{t_0 < t_i \leq t} v_i
\end{equation}
with $v_i$ the value of a transaction occurring at time $t_i$,
and $t_0$ some arbitrary reference time.
The definitions for $N$ and $D$ are similar, respectively counting the number of ticks
and transactions between $t_0$ and $t$.
Different measures of activity are given by finite differences of these
integrated quantities.
The (mean) transactions value $dV/dt$, in an interval $\Delta t$,
measure in GBP per hour, is defined by
\begin{equation}
    \frac{dV}{dt}(t) = \frac{V(t_0, t) - V(t_0, t - \Delta t)}{\Delta t/1h}
    = \frac{1h}{\Delta t}\sum_{t - \Delta t < t_i \leq t} v_i
\end{equation}
where $1h$ means 1 hour (the ratio of two time intervals, like $\Delta t/1h$, is a dimentionless number).
The (mean) transaction value per transaction $dV/dD$, in an interval $\Delta t$, is defined by
\begin{equation}
    \frac{dV}{dD}(t) = \frac{V(t_0, t) - V(t_0, t - \Delta t)}{D(t_0, t) - D(t_0, t - \Delta t)}.
\end{equation}
The (mean) number of transaction per tick $dD/dN$,
in an interval $\Delta t$, is defined as
\begin{equation}
    \frac{dD}{dN}(t) = \frac{D(t_0, t) - D(t_0, t - \Delta t)}{N(t_0, t) - N(t_0, t - \Delta t)}.
\end{equation}
The (mean) tick rate $dN/dt$ in an interval $\Delta t$, measured in tick per hour,
is defined by
\begin{equation}
    \frac{dN}{dt}(t) = \frac{N(t_0, t) - N(t_0, t - \Delta t)}{\Delta t/1h}.
\end{equation}
The exact definitions of the activities measured with the order queue
and with the price dynamic are given respectively in section
\ref{sec:orderQueueScaling} and \ref{sec:pricesScaling}.

We use 16.5 months of data for the companies included in
the FTSE 100 in order to make a cross-sectional study
of the dependency of the ``activities'' with respect to the ``size'' of the companies.
These companies have a market capitalization between 1 to 100 billion GBP,
and therefore cover roughly 2 orders of magnitude.
The main empirical findings are that there are clear scaling relations between
the company size and the measure of ``activities'':
\begin{enumerate}
\item
The tick rate $dN/dt$ scales with the market capitalization $M$ with an exponent 0.39.

\item
The value per transaction $dV/dD$ scales with the market capitalization $M$ with an exponent 0.44.

\item
The number of transactions per tick $dD/dN$ is roughly constant,
with an average of 1 transaction every 5 to 10 ticks.

\item
The transaction value per hour $dV/dt$ scales as the the market capitalization $M$
with the exponent 0.90 (whereas a linear dependency could be expected).

\item
The value in the order queue $v_\text{queue}$ increases with the the market capitalization $M$
with the exponent 0.76, markedly below 1.

\item
The tick-by-tick volatility $\tickVol$ decreases with the capitalization $M$, with an exponent -0.23.

\item
The annualized volatility $\annVol$ is independent of the market capitalization $M$,
contrary to the belief that smaller capitalizations have larger volatility.

\item
The spread $S$ increases with the tick-by-tick volatility $\tickVol$, with an exponent 0.94.

\end{enumerate}
None of these relations could be anticipated,
most notably relations 4 and 5 where a simple linear scaling could be expected.
As the author is not aware of any market microstructure theory explaining these relations,
this paper is mostly empirical.
On the theoretical side, a simple argument about the independence of averages
allows to relate the exponents 1, 2, 3 and 4.
In section \ref{sec:pricesScaling}, a random walk argument is used to
relate the exponents in point 1, 6 and 7.
In the last section, using the scaling relations 2 and 4, combined with a random walk argument,
we derive the exponent 6.
Yet, the other values for the exponents beg for an explanation.

Scaling relations are observed in many areas in natural sciences.
They are convenient because a scaling law captures in a simple
formula the dependency between two quantities.
Yet, the major importance of scaling laws originates in the underlying
mechanism needed to create them:
for a scaling dependency to appear over a broad range of parameters,
a robust mechanism is needed to erase small differences.
A historical example is the second order phase transitions,
which show accurate scaling over several orders of magnitude,
and with the same exponents for very different physical systems.
A diverging correlation length at the phase transition is the mechanism that
erases the differences between systems,
and this understanding lead to the development of
the renormalization group and critical phenomenon in physics.
A more formal analogy between financial market  and critical phenomenon in physics
has been proposed recently
\cite[and references therein]{marsili.2003}.
In this paper, the author argues that for one asset, in the time series direction,
many stylized facts are resulting from a critical behavior induced by the
trading mechanisms.
Similarly, while considering correlations across many stocks,
the emerging structure is again scale free and reminiscent of
critical  phenomenon.

In this critical phenomenon perspective,
a scaling law is a signature for a strong underlying
mechanism that need to be understood.
The scaling laws are giving important informations
about the behavior of the microscopic system,
in our particular case the orders and trades made by the market agents.
For the present study, the mechanism at work must originate in
the behavior of the market participants, in their appetite for profit and risk,
and in the trading strategies they use to achieve their objectives \cite{treynor.1971}.
Yet, the connection between the behavior of the market's agents
(microscopic) and the resulting price
and ``activity'' statistical properties (macroscopic)
remain to be understood.
For a given microstructure market theory,
the possible explanation of the observed scaling laws
is a very strong selection criterion.
A recent example \cite{gabaix.2003.1, gabaix.2003.2}
of such theory in finance explains why the probability density
of the return decays as $p(r) \sim |r|^{-4}$ for large return,
for a very broad class of assets:
the underlying mechanism that creates this scaling law is the ``price impact function''
that grows as the square root of the transaction size.
Another scaling study on the relationship between volatility and
company size can be found in \cite{plerou.1999, plerou.2000},
and its relation to the present work is discussed in sec.~\ref{sec:pricesScaling}.

We analyze the scaling relations using data over the period
1.8.2001 to 14.12.2002.
This is quite an eventful period for the stock market,
with a large volatility and a global downward trend.
It includes in particular the September 11, 2001 attack.
The dispersion is also enhanced by
the different activity sectors for the companies entering the FTSE 100,
like airline transportation, high tech, retail, banking, or pharmaceutical,
and we could expect large deviations around the observed scaling.
Beside, a few companies experienced extraordinary events.
The largest outlier is the company PowerGen (code PWG)
which was bought by the German company e.on at the end of the second quarter 2002,
and was removed from the FTSE on 2.7.2002.
Clearly, this kind of events are outside the normal trading activity,
and only this single company was removed from our data set.
Despite the peculiarities of the time period and diversity of the companies,
most of the scaling relations are very well observed.
In order to estimate quantitatively the scaling exponents,
we used a least square procedure,
but with an error for both variables because the capitalization is also a fluctuating quantity.
This more complex procedure correctly takes into account
the fluctuations in measuring both the market capitalization and the ``activity''.

The plan of this paper is as follows.
The data set is presented in the next section,
followed by the technical part on computing the standard deviations and least square estimates.
The measure of the company ``size'' with the market capitalization or
with the number of shares is discussed in sec.~\ref{sec:companySize}.
In sections \ref{sec:transactionScaling}, \ref{sec:orderQueueScaling} and \ref{sec:pricesScaling},
we investigate respectively the scaling for quantities
related to the trades, for the order queue and for the prices dynamic (spread, volatilities).
All the results are discussed in the conclusion, 
where an argument is given to explain the scaling for the tick-by-tick volatility.
All the values for the exponents are collected in 2 tables included in this last section.

\section{The data}
The data set covers the FTSE 100 stocks, from 1.8.2001 to 14.12.2002.
The ticks are obtained from Reuters, and are called ``level 1'' and ``level 2'' data.
For both streams, each tick contains a time stamp
recorded by the computer storing the data.
The level 1 data are meant to drive the tickers on the trader screens.
They contain essentially the best bid, ask, 
cumulative numbers of transactions and cumulative transaction volume.
The volumes in the queue at the best bid and ask are not given 
(the fields are in the data structure, but with a zero value).
The relationship between the events in the market (new order, transaction, withdraw, ...)
and the level 1 ticks are fairly complex.
A simple transaction is reflected in two ticks, one that updates the best prices,
one that updates the cumulative transaction count and volume.
In general, one level 1 tick is related to an ``atomic'' cause,
like update of the best prices, or update of the transaction count and volume.
A (large) transaction that matches several orders in the queue results in more ticks.
The cause for each tick is not given in the data stream, and a lower best bid or higher best ask
can be due either to a transaction or to a withdrawn order.
This ambiguity could be resolved by using the neighboring ticks,
but to make the matter worst, the tick due to the price
adjustment of a transaction might come before or after the tick(s) for the
adjustment of the transaction count and cumulative volume.
All these features make detailed computations from level 1 data cumbersome.
In particular, the tick rate is a bad measure for the traders activity,
and it is not possible to compute withdrawn volumes
(for example, withdraws leaving the best bid or ask
unchanged are not reflected in level 1 data).

In the level 2 data, each tick contains the time stamp
and the first 10 positions on the bid and on the ask sides of the order queue.
For each successive positions are given the price, number of stocks and number of orders.
A new tick is issued for every change in the (first 10 positions of the) order queue,
which can be a new order, a transaction or a canceled order.
In the level 2 ticks, no further information is provided,
and in particular the cause for a new tick is not given (new order, transaction, withdraw).
Yet, these data provide for an accurate measure of the market activity,
as they contains the flow of incoming orders.

In principle, all the data needed for the computations presented
in this paper can be extracted separately from level 1 or from level 2 data.
Yet, with an independent extractions, the withdrawn orders remain unknown.
Because it was not clear {\it a priori} what is the importance of withdrawn orders,
we decided to match both streams.
By matching the data streams, the cause of each tick can be infered,
in particular a decrease at the best bid or ask can be attributed
to a transaction or to a withdrawn order.
The analysis of withdrawn orders shows a constant small fraction of them per tick,
without incidence on the scaling.
Therefore, the results on withdrawn orders are not presented in the paper.

The matching of both streams is not very easy,
because the time stamps are given by the computer that captures the data,
and a random lag of up to 6 seconds can be observed between level 1 and 2 data streams.
The strategy to match both streams is fairly pedestrian: on the level 1 data stream,
the tick corresponding to the transactions are identified.
On the level 2 data stream, when the number of orders at the best bid or ask decreases,
this can be due either to a transaction or a withdrawn order.
This ambiguity is resolved using level 1 data occurring within 6 seconds of the level 2 tick,
by searching for a trade of the same number of shares and price.
In this way, the atomic cause of each tick can be inferred,
and consistent values for all relevant quantities can be computed
using jointly both data streams.

As both data streams originate from an electronic trading platform,
it might be expected that they are error free.
This is {\em not} the case.
For example the relation bid $<$ ask can be broken for both data streams.
On level 2 data, the prices are expected to be ordered by increasing values in the
order queue, but a few ticks are not correctly ordered.
Both examples are quite shocking,
and we do not know the cause for the breakdown of such basic relationships.
For these reasons, the data have been filtered,
with a typical rejection rate below one per thousand.

The capture of the data suffers from a few breakdowns,
resulting in incomplete or missing data for these days.
For the companies that existed through the whole sample,
we have a maximum of 328 full trading days.
Then, a number of companies entered or exited the FTSE100 during that period.
We have excluded from the scaling analysis the stocks with less than 200 trading days.
Finally, the company PWG, traded for 207 days,
appears as a clear outlier for the reason mentioned in the introduction,
and was removed from the sample.
Similarly, the stock SDRt (the non voting share for Schroders) has been removed.
It is included in the FTSE because coupled with the stock SDR,
but its  mean capitalization of $0.46 ~10^9$ GBP
is quite below the lower limit of the FTSE at around $10^9$ GBP.
The stock SDR, with a mean capitalization of $1.2~10^9$ GBP is included in the data sample.
All the quantitative results presented in this paper
are robust with respect to the exclusion of these two stocks,
or with respect to the choice of the minimum number of days.
The final sample contains 92 companies.

\section{Least square estimates}
\label{sec:regression}
Our goal is to search for scaling behavior depending on the
size of the companies, mainly in form of power laws.
By taking logarithm of the quantities under investigation,
a scaling behavior is equivalent to search for linear dependencies,
and the best fit can be obtained with a least square (LSQ) estimate.
Yet, in order to compute meaningful least square estimates,
the standard deviation must also be available.
We proceed in the following way to compute the mean and the standard deviation
for each stock.
For each quantities of interest, the daily mean is computed,
with the daily mean being a sum over all the ticks,
divided by the number of ticks for this day.
We obtain in this way a time series for each quantity and each stock, sampled daily.
Then, say for a daily quantity $z_k$, we compute the mean $\bar{z} = E z_k$,
absolute mean $\overline{|z|} = E |z_k|$
standard deviation $\sigma_2^2 = E (z_k - \bar{z} )^2$,
and absolute standard deviation $\sigma_1 = E |z_k - \overline{|z|} |$
of these daily time series, where $k$ is a daily index.
The kurtosis is estimated by the ratio $\kappa = \sigma_2/\sigma_1$.
For all quantities studied in this paper, the distributions are leptokurtic with $\kappa > 1$,
pointing to distributions that are more fat tailed than a Gaussian distribution.
For a given quantity, the average over the stocks of the kurtosis is between
1.1 (for the number of transaction per tick)
to 1.6 (for the L2 tick-by-tick volatility).
Notice that this corresponds to fairly strongly leptokurtic distributions.
The mean and absolute standard deviation for each stock are used as inputs in LSQ computations.
For a pair of quantities $x$ and $y$, we have the absolute standard deviations in both variables.
The usual LSQ estimate assumes a variance in the $y$ variable,
but none in the $x$ variable which is assumed to be known ``exactly''.
Yet, the capitalization is a time series, and therefore has its own variance.
Therefore, we have to use a more sophisticated LSQ estimate,
with an error in both variables.
We have used the procedure described in
\cite{NumRecipes} to carry out the computations.
Notice that this LSQ estimate is more complex to compute as
it  involves a minimization problem
(to find the best parameters) and to find the roots of a one dimensional
function (to compute the error on the parameters).
The value of the minimum (slope and intercept)
and the errors on the parameters (standard deviations of the slope and intercept )
are fairly insensitive to the choice
for the standard deviation ($\sigma_1$ versus $\sigma_2$),
but the goodness of fit depends directly on the choice for the standard deviation.
As $\sigma_1$ is lower than $\sigma_2$, this produces systematically worst goodness of fit.
For this reason, we have used the more conservative absolute standard deviation $\sigma_1$.

\section{Measuring the size of companies}
\label{sec:companySize}
Our goal is to see how quantities related to the trading of
stocks change with the size of companies.
Therefore, the first question is how to measure the ``size'' of traded companies.
The obvious candidate is the market capitalization,
but another possible measure is the number of shares.
The number of shares seems less plausible, because we expect a kind of  ``gauge invariance''
between the price and the number of shares:
if the number of shares is multiplied by a number and the stock price divided by the same number,
all transaction values and total market capitalization remain invariant.
However, what has economic value is the number
of monetary unit exchanged (the number of GBP),
and not the number of shares.
Indeed, companies with stock at a high price per share may undergoes a stock splitting,
while keeping the market capitalization invariant during the operation
(this stock splitting is the analogous of a "gauge transfomation").
Therefore, the pertinent quantities are of the form price times number of shares $p n$,
as in a transaction value or the total market capitalization,
and not separately the price $p$ or number of shares $n$.

\begin{figure}[htb]
  \centering
  \includegraphics[width=0.8\textwidth]{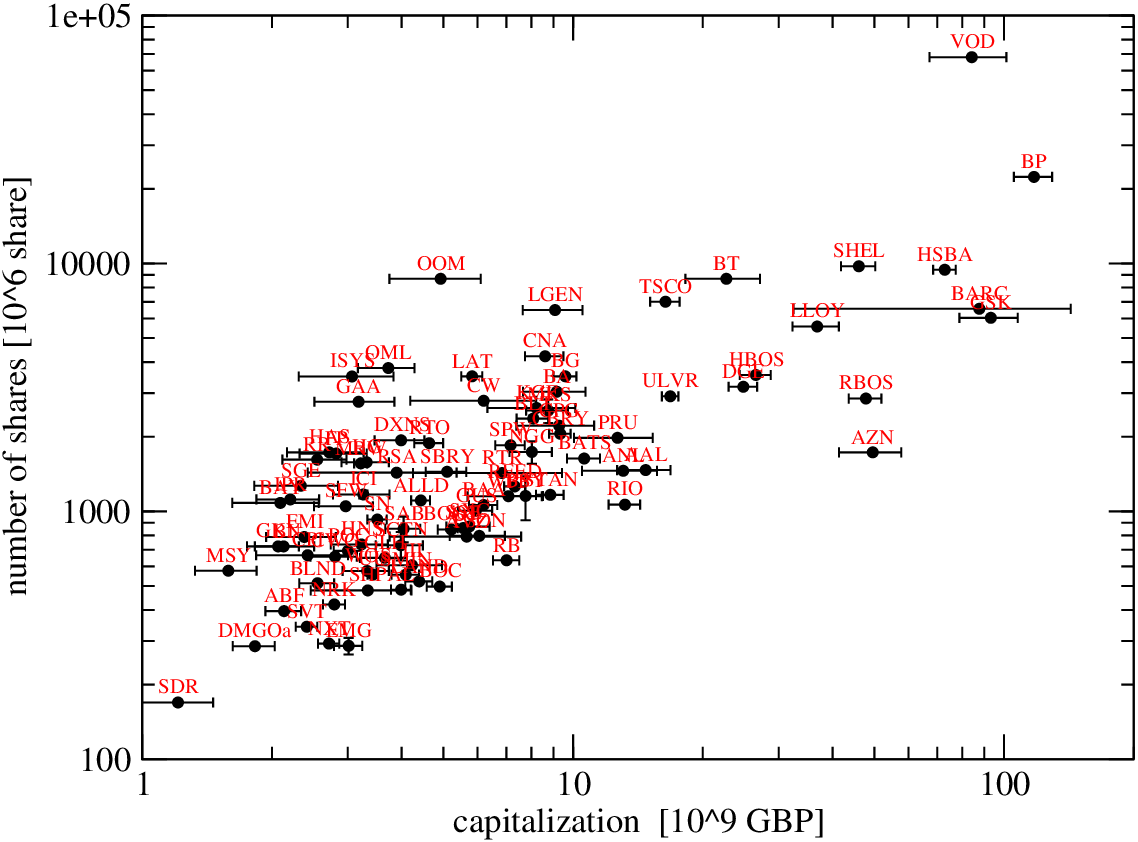}
  \caption{\small \sf The number of shares versus the mean capitalization $M$.
  For most stocks, the number of shares is unchanged in the data sample,
  and therefore the corresponding variance is zero.
  }
  \label{fig:nbShareVsCap}
\end{figure}

The ``gauge invariance'' is a theoretical argument in
favor of using the market capitalization,
but the situation is not so clear cut on the empirical side.
As shown in Fig.~\ref{fig:nbShareVsCap}, the number of shares is
roughly proportional to the total market capitalization.
Indeed, most stock prices are between 1 and 10 GBP, without any clear dependency
between prices and capitalization.
This implies the near proportionality between market capitalization and number of shares,
but a least square estimate with these two quantities is strongly rejected.
Therefore, both the capitalization and the number of shares could be
used as similar but different measure of  ``size'' for the companies.

For all the empirical scaling relations presented afterward,
the market capitalization has systematically a better goodness of fit than
the number of shares, in agreement with the ``gauge invariance'' argument.
As it has more explanatory power,
we will present the figures below only for the market capitalization as a
measure of the company sizes,
but we have checked systematically the results with the number of shares,
both at the graphical level and with the quantitative least square estimates.
To summarize the whole issue, the near proportionality of market capitalization and number of shares
makes the investigation of the relevant measure of ``size'' a bit subtle.
Beyond the theoretical ``gauge invariance'' argument, it is only the goodness
of fit of alternate estimates made with the market capitalization or with the number of shares
that provide the decisive criterion in favor of the first one.

Finally, the mean market capitalization can be computed with an arithmetic or a geometric mean:
\begin{eqnarray}
   M_\text{arithmetic} & = & \frac{1}{n} \sum M_i   \\
   M_\text{geometric} & = & \exp\left(\frac{1}{n} \sum \ln(M_i) \right)   \nonumber
\end{eqnarray}
where $M_i$ is the capitalization for tick $i$ and $n$ is the total number of ticks.
Both quantities differ by less than 1\% for most stocks, and the largest difference is less than 7\%.
Moreover, the choice of formula for computing the mean capitalization
has essentially no influence on the scaling results below.
In order to avoid too large an influence of the large stock prices in the average,
we have used the geometric mean.

\section{Empirical scaling analysis for the transactions}
\label{sec:transactionScaling}
\begin{figure}[htb]
  \centering
  \includegraphics[width=0.8\textwidth]{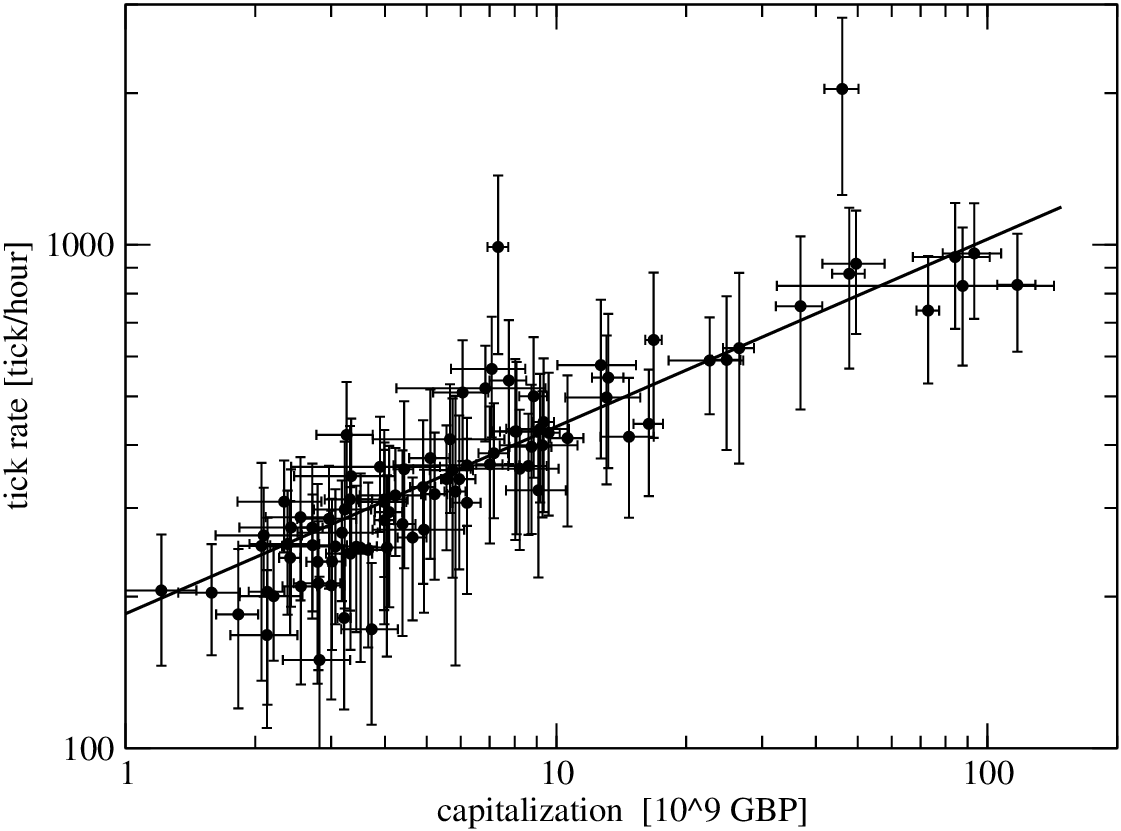}
  \caption{\small \sf The tick rate $dN/dt$ versus the capitalization $M$.
  The continuous line corresponds to the $dN/dt = 180 ~M^{0.39}$
  }
  \label{fig:tickRateVsCap}
\end{figure}

\begin{figure}[htb]
  \centering
  \includegraphics[width=0.8\textwidth]{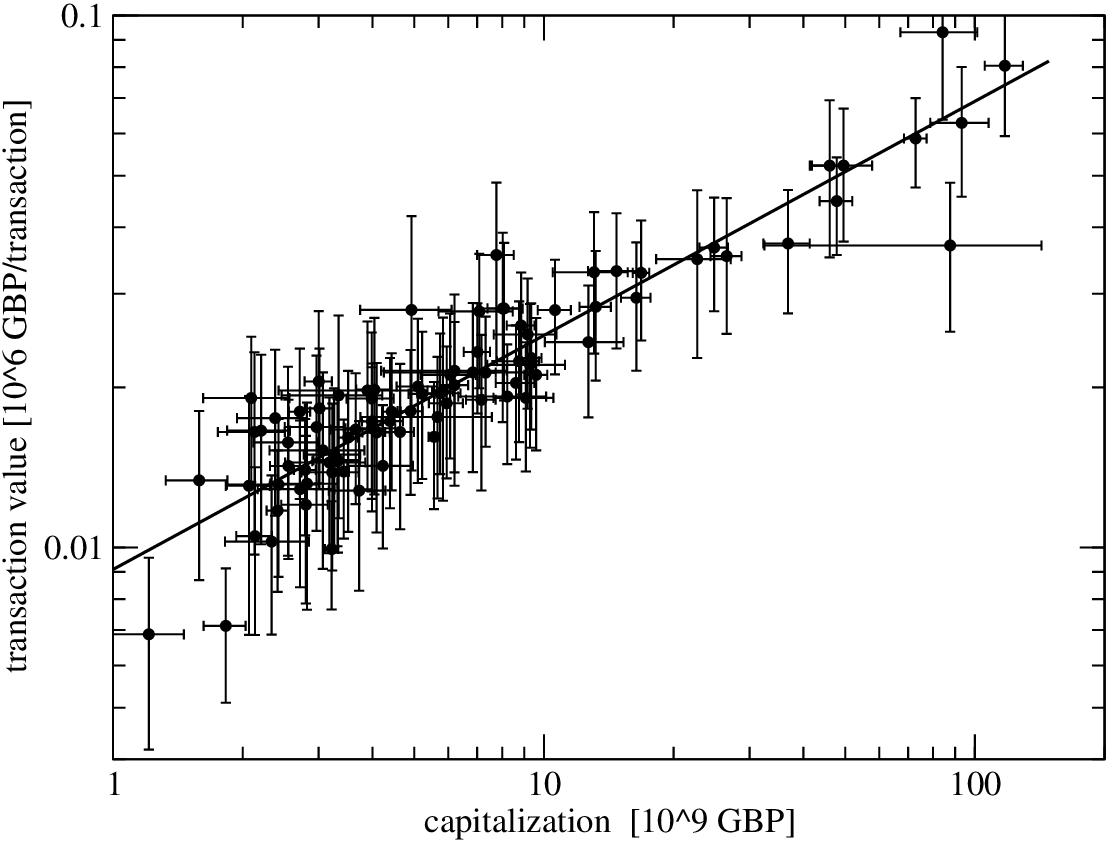}
  \caption{\small \sf The transaction value per transaction $dV/dD$ versus the capitalization $M$.
  The continuous line corresponds to $dV/dD = 9'100 ~M^{0.44}$, with the transaction value in GBP
  and the capitalization given in $10^9$ GBP.
  }
  \label{fig:transactionValueVsCap}
\end{figure}

\begin{figure}[htb]
  \centering
  \includegraphics[width=0.8\textwidth]{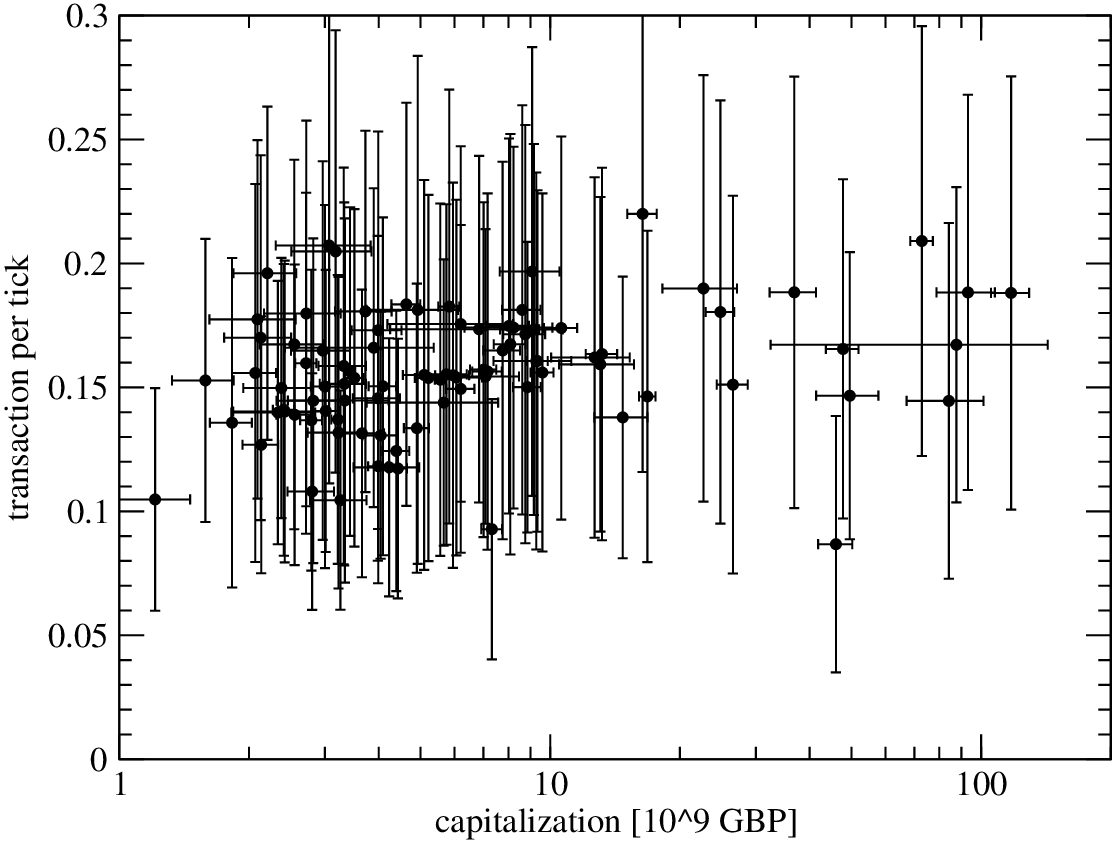}
  \caption{\small \sf The mean number of transaction per tick $dD/dN$ versus the capitalization $M$.
  }
  \label{fig:transactionPerTickVsCap}
\end{figure}

In this section, we focus on the scaling relation between the market capitalization
and quantities related to the order flow and transactions.
The first interesting relation is the scaling between the
tick rate and the capitalization,
as given in Fig.~\ref{fig:tickRateVsCap}.
The LSQ estimate indicates that the tick rate scales with the capitalization,
with an exponent 0.39 $\pm$ 0.03.
This is markedly below a square root dependency,
and excludes the exponent 0.5 at a 4 $\sigma$ level.
The relation between the transaction value per transaction and the capitalization
is shown in Fig.~\ref{fig:transactionValueVsCap}.
Again, a clear scaling relationship appears, with an exponent 0.44 $\pm$ 0.03.
The transaction value per tick $dV/dN$
(equivalent to the total transaction value divided by the total number of ticks)
has an exponent 0.48 $\pm$ 0.06, compatible with a square root dependency.
In Fig.~\ref{fig:transactionPerTickVsCap}, the number of transaction per tick is shown
as a function of the  capitalization.
In a first approximation, regardless of the capitalization,
there is in average 1 transaction every 5 to 10 ticks.
The exponent obtained by LSQ estimate is 0.05 $\pm$ 0.05, and is compatible with zero.

\begin{figure}[htb]
  \centering
  \includegraphics[width=0.8\textwidth]{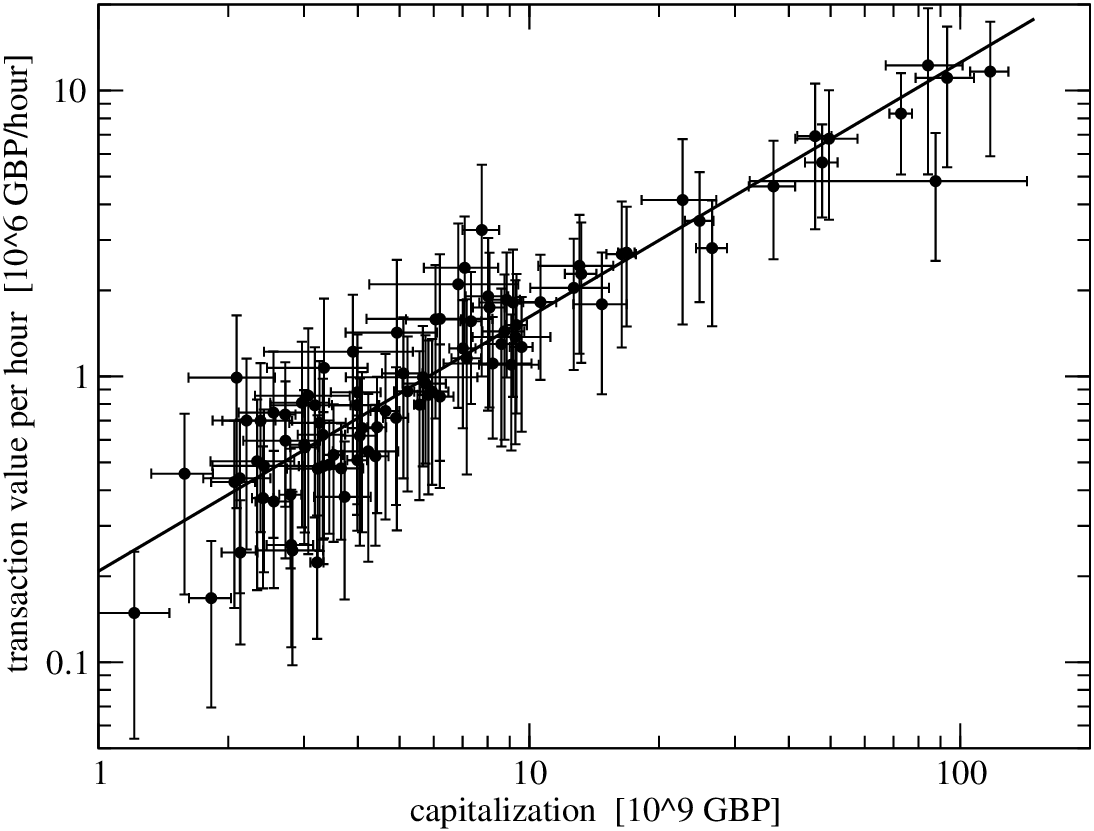}
  \caption{\small \sf The mean transaction value per hour $dV/dt$ versus the mean capitalization $M$.
  The continuous line corresponds to $dV/dt = 0.208 10^{-3} M^{0.90}$
  (both measured in the same monetary unit).
  }
  \label{fig:transactionRatePerHourVsCap}
\end{figure}

Finally, we can analyze the transaction value per hour $dV/dt$.
The near proportionality with the capitalization can be checked on
Fig.~\ref{fig:transactionRatePerHourVsCap}.
Yet, the least square estimate gives an exponent of 0.90 $\pm$ 0.055,
namely is below the value 1 at a 2$\sigma$ level.
This result is robust against change in the minimum number of days,
exclusion of outliers, or the definition of the variance that is used
(with absolute value or square), namely the value 1 is always rejected at a 2$\sigma$ level.
This result is surprising, because intuitively, a simple proportionality is expected.
An argument about short term arbitrageur says that they should prefer very liquid stocks
as the spread is smaller, and therefore the transaction cost lower.
This argument leads to an exponent larger than 1, contrarily to what we observe.
Another argument says that enough volatility is needed to trade on a short term basis.
As the volatility is expected to decrease with increasing capitalization,
this would lead to an exponent smaller than 1.
Yet, the empirical relationship between annualized volatility
and capitalization shows an exponent compatible with zero (see sec.~\ref{sec:pricesScaling}),
and therefore this argument cannot be used.
Even though a simple proportionality is expected between
transaction value per hour and capitalization,
it is not clear why this should hold, and many mechanisms can break the simple proportionality.
Indeed, an exponent smaller than one is observed empirically in this data set.

From their respective definitions, the transaction value per hour is given by the product
\begin{eqnarray}
         \frac{dV}{dt} \left[\frac{\mbox{GBP}}{\mbox{hour}}\right] & = &
         \frac{dV}{dD} \left[\frac{\mbox{GBP}}{\mbox{transaction}}\right]
  ~\cdot~\frac{dD}{dN} \left[\frac{\mbox{transaction}}{\mbox{tick}}\right]  \nonumber
  ~\cdot~\frac{dN}{dt} \left[\frac{\mbox{tick}}{\mbox{hour}}\right].
  \label{eq:scalingRelation}
\end{eqnarray}
Taking expectations and assuming independence in the right hand side,
then inserting the above scaling relations, we obtain a simple sum for the exponents.
Inserting the exponents computed above for terms in the right hand side leads to the value 0.86,
within the error margin of the exponent 0.90 computed for the left hand side.

\section{Empirical scaling analysis for the order queue}
\label{sec:orderQueueScaling}
In this section, we analyze the scaling with the market capitalization
of the ``activity'' related to the order queue.
In order to characterize the order queue at tick $i$,
two quantities measuring its activity are interesting,
namely the total number of orders $n_{\text{queue}, i}$ and the total value $v_{\text{queue}, i}$.
Yet, we would like to measure the activity of the queue around the mid-price,
and cut-off orders far away as they are irrelevant for the short term price dynamic.
For this purpose, we introduce an exponential cut-off
to compute the mean number of orders and mean value
\begin{eqnarray}
   n_{\text{queue}, i}(\mu) & = &
   	\sum_{q\in\text{order queue}} n_{i,q} ~\frac{\exp(-10000|x_{i,q} - x_i|/\mu)}{\mu}  \\
   v_{\text{queue}, i}(\mu) & = &
   	\sum_{q\in\text{order queue}} v_{i,q} ~\frac{\exp(-10000|x_{i,q} - x_i|/\mu)}{\mu} \nonumber
\end{eqnarray}
with $x_i$ the (prevailing) logarithmic middle price (see next section),
and for each position $q$ in the queue,
$n_{i,q}$, $v_{i,q}$, and $x_{i,q}$ denote respectively
the number of orders, the value, and the logarithmic price.
The sum over the order queue is exponentially cut-off with the distance from the logarithmic mid-price,
and the parameter $\mu$, measured in basis point (1 BIPS = $10^{-4}$), controls the effective depth
of the order queue used to compute the average.
For the scaling analysis,
we computed the daily average $n_\text{queue}$ and $v_\text{queue}$
from their respective tick-by-tick quantities.

The value of the parameter $\mu$ that should be used depends on several considerations.
First, the smallest price increment that can be used for a given stock
is set by the London Stock Exchange (LSE),
and depends from the actual price.
With $p$ the price and $\Delta p$ the price increment (both in GBP),
the exact rule is as follows
\begin{equation}  \label{eq:minPriceIncrement}
	\Delta p = \begin{cases}
	0.0025 \hspace{3em} & \text{for } p < 5        \\
	0.005               & \text{for } 5 < p < 10   \\
	0.01                & \text{for } 10 < p
	\end{cases}
\end{equation}
For most actual prices per share,
the smallest price increment is between 5 to 10 BIPS,
as can be see on fig.\ref{fig:spreadVsPrice}.
A second factor influencing the choice of $\mu$ is the typical queue shape.
A computation of the distribution for the mean number of orders and values
as a function of the distance to
the middle price gives distributions that are in agreement
with the results of \cite{BMP.2002-03} for small distance to the mid-price
(we cannot compare for large distance as only the first 10 positions in
the order queue are available on the LSE).
In \cite{BMP.2002-03}, the authors use data from the Paris stock exchange
from the incoming orders and trades to reconstruct the full order book.
They also derive the distributions of order and value analytically with a ``zero intelligence model'',
where the price follows a random walk and annihilates orders on its path.
The empirical distributions are symmetric around the middle price,
with a maxima at 3 to 4 price increments away from the middle price,
pointing to a small value for $\mu$.
The last factor influencing the choice of $\mu$ is that the level 2 data
contain the first 10 positions on each side, and the queue further out is unknown.
This factor limits $\mu$ on the upper side.
Together, these 3 requirements on $\mu$ lead to {\it a priori} values between 4 to 40 BIPS.

The study of $n_\text{queue}(\mu)$ and $v_\text{queue}(\mu)$ as a function of
the parameter $\mu$ shows a smooth dependency,
with a maxima between $\mu =$ 10 to 20 BIPS for all stocks.
These values are in agreement with the estimations in the previous paragraph
(a small value for $\mu$ would pick only the best bid and ask orders,
whereas a large value corresponds to a simple sum over the queue).
Therefore, we have computed the mean number of orders
and queue value using $\mu =$ 10 BIPS for all stocks,
and have checked that the results below do not depend on this value.

\begin{figure}[htb]
  \centering
  \includegraphics[width=0.8\textwidth]{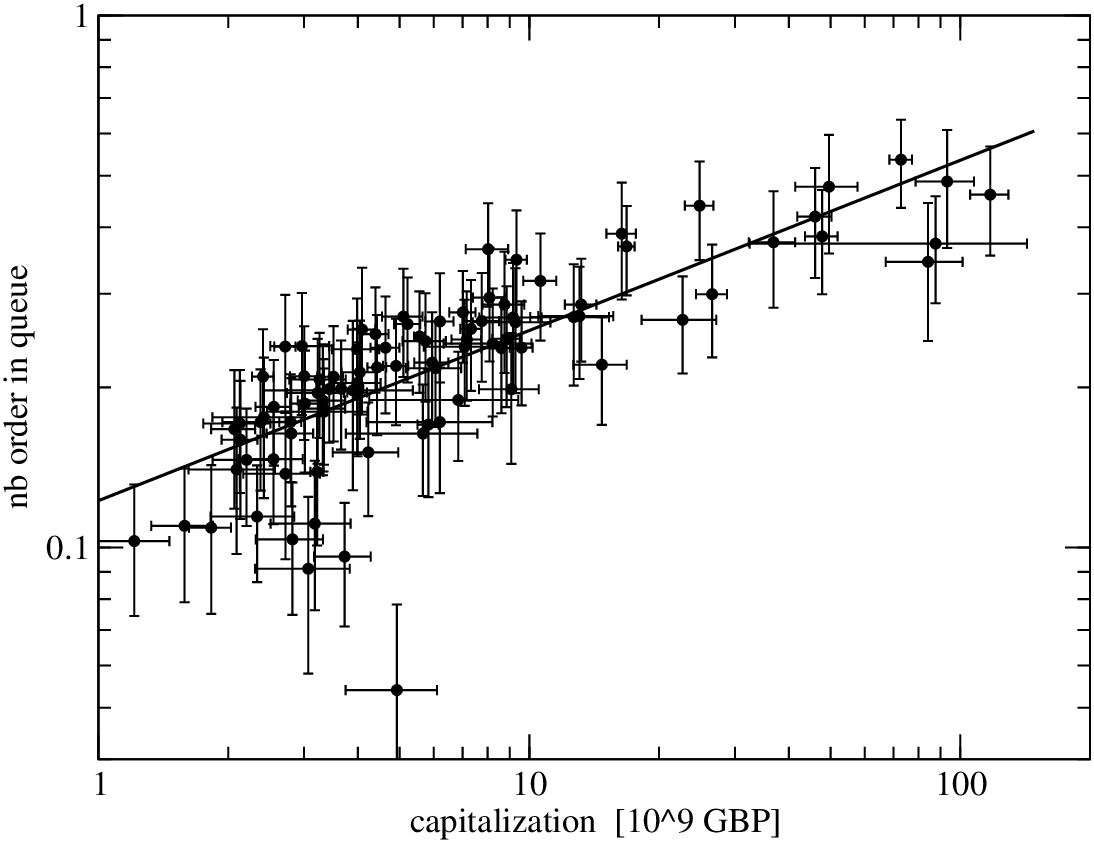}
  \caption{\small \sf The number of orders in the queue $n_\text{queue}$ versus the capitalization $M$.
   The continuous line corresponds to $n_\text{queue} = 0.12 (M)^{0.32}$
  }
  \label{fig:nbOrderInQueueVsCap}
\end{figure}
\begin{figure}[htb]
  \centering
  \includegraphics[width=0.8\textwidth]{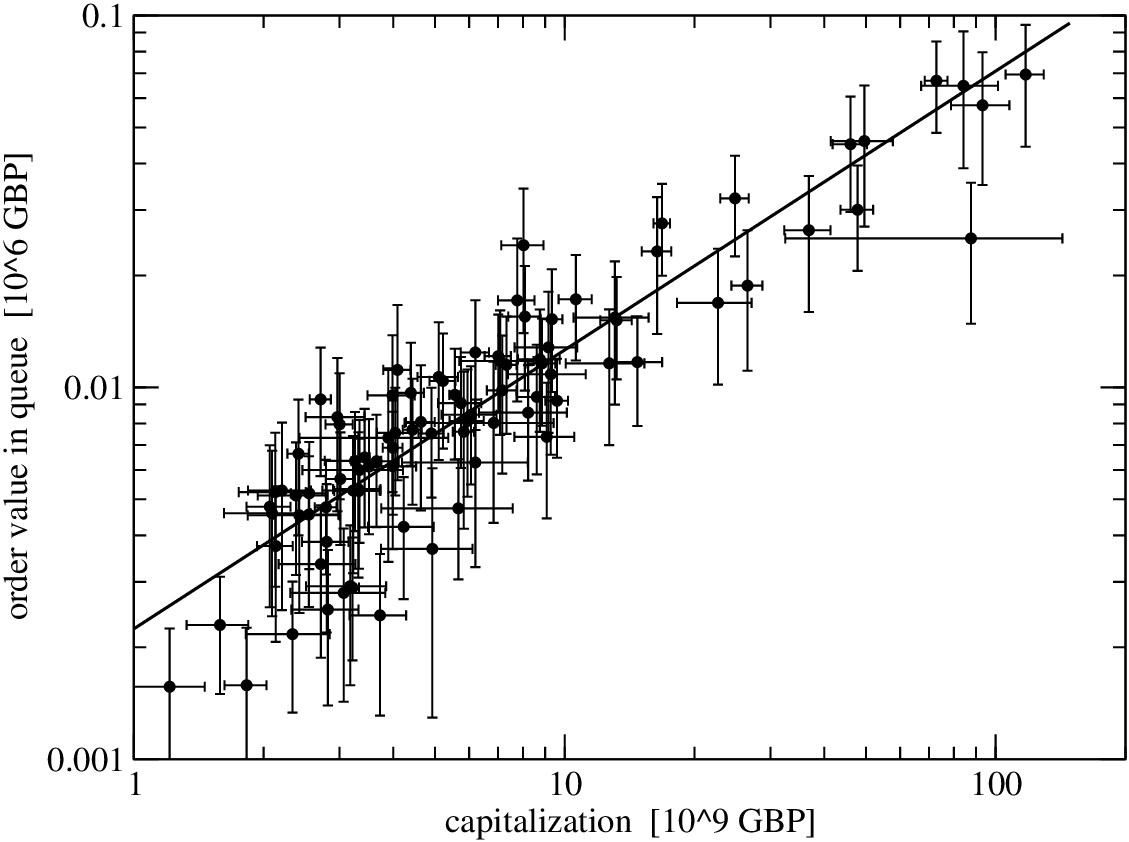}
  \caption{\small \sf The value in the queue $v_\text{queue}$ versus the  capitalization $M$.
   The continuous line corresponds to the $v_\text{queue} = 2.2 10^{-6} (M)^{0.76}$,
   with both values measured in the same unit.
  }
  \label{fig:valueInQueueVsCap}
\end{figure}
The number of orders in the queue is plotted in Fig.~\ref{fig:nbOrderInQueueVsCap},
with a scaling exponent of 0.32 $\pm$ 0.03.
The value in the order queue is plotted in Fig.~\ref{fig:valueInQueueVsCap},
and the exponent is 0.76 $\pm$ 0.04.
In particular, this last exponent value excludes an exponent of 1 at a 6$\sigma$ level,
whereas we could have expected a simple linear dependency between
market capitalization and value in the queue.

With the hypothesis that the exponent 0.44 for the transaction value per transaction $dV/dD$ is identical
to the exponent for the value per order in the queue,
and using the decomposition that the value in the queue is the number
of order in the queue times the value per order,
we obtain the exponent 0.32 + 0.44 = 0.76, in agreement with the exponent 0.75.
Both scaling relations for the queue are roughly in line with the scaling for the transactions.
The number of orders and tick rate have exponents 0.32 and 0.39 respectively, in fair agreement.
However, the value in the queue and transaction per hour have respectively exponents 0.75 and 0.90,
in marginal agreement, but both below 1.

\begin{figure}[htb]
  \centering
  \includegraphics[width=0.8\textwidth]{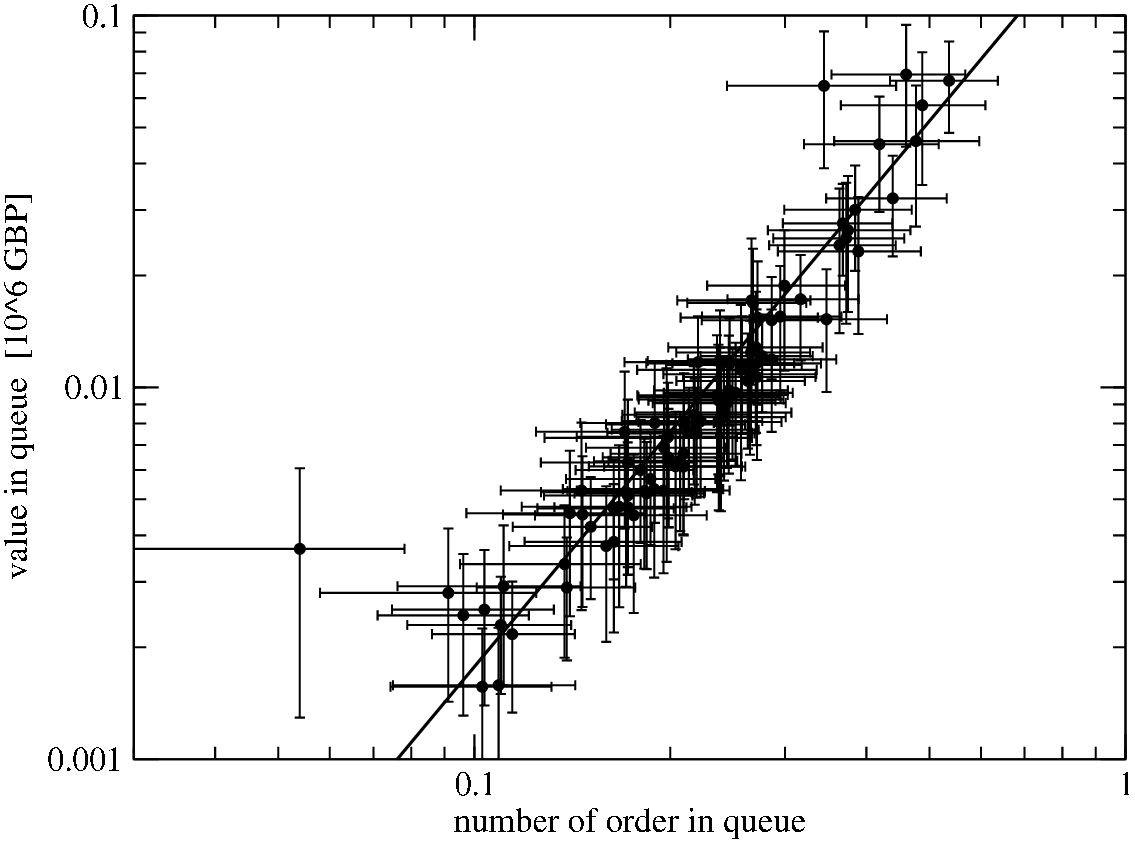}
  \caption{\small \sf The value in the order queue $v_\text{queue}$
  versus the number of orders in the queue $n_\text{queue}$.
   The continuous line corresponds to $v_\text{queue} = 0.22 ~ n_\text{queue}^{2.1}$
  }
  \label{fig:valInQueueVsNbOrderInQueue}
\end{figure}

\begin{figure}[htb]
  \centering
  \includegraphics[width=0.8\textwidth]{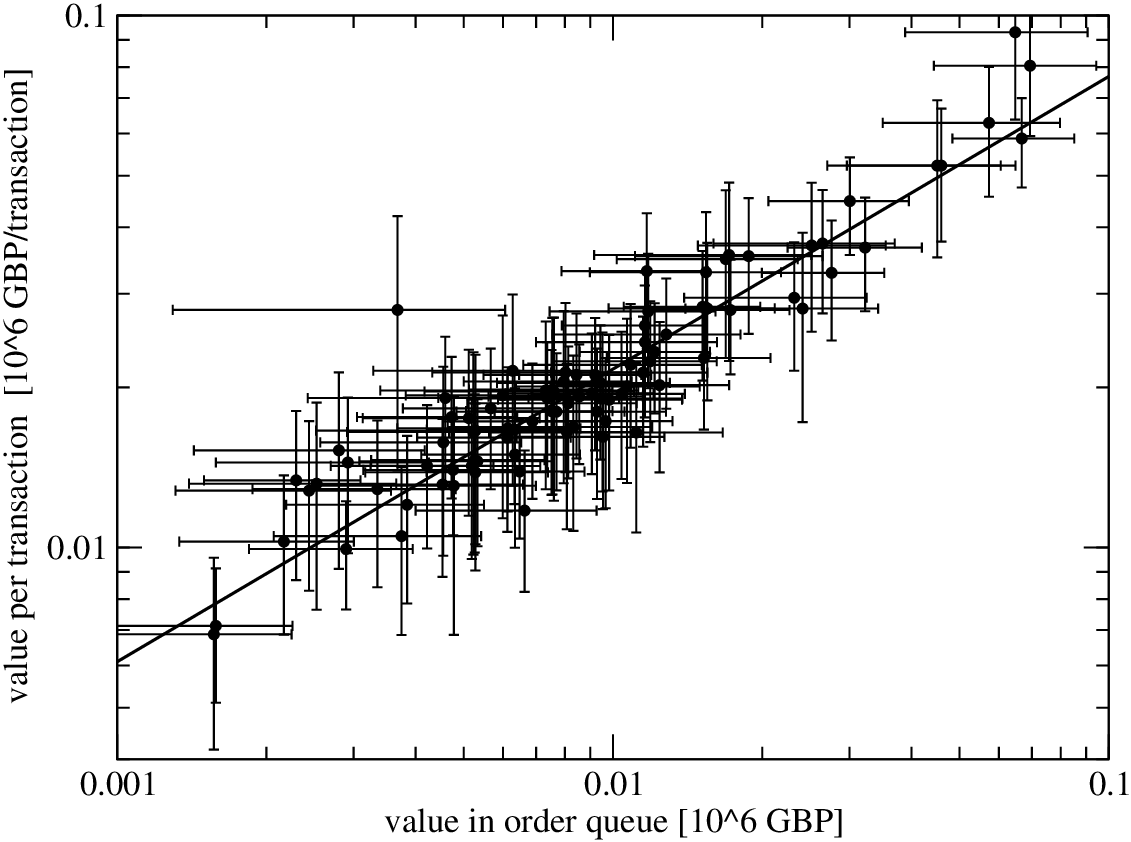}
  \caption{\small \sf The value per transaction $dV/dD$ versus the value in the order queue $v_\text{queue}$.
   The continuous line corresponds to the $dV/dD = 0.30 ~(\text{value in queue})^{0.57}$
  }
  \label{fig:valuePerTransactionVsValueInQueue}
\end{figure}

A very clear scaling relation is obtained for the value in the queue as a function
of the number of orders in the queue, as shown on Fig.~\ref{fig:valInQueueVsNbOrderInQueue},
with an exponent 2.1 $\pm$ 0.2.
Finally, the transactions can be related to the queue.
As shown on Fig.~\ref{fig:valuePerTransactionVsValueInQueue},
the value per transaction scales with the value
in the order queue with an exponent 0.57 $\pm$ 0.05.
This value is in agreement with the above scaling relations $dV/dD \sim M^{0.44}$
and $v_\text{queue} \sim M^{0.76}$,  leading to an
exponent 0.58 when eliminating $M$ between the two relations.

\section{Empirical scaling analysis related to the price dynamics}
\label{sec:pricesScaling}
In the previous two sections, we have studied the transactions and the order queue.
Their dynamical interactions generate the best bid and ask prices as a function of time.
For this cross-sectional study of scaling relationships,
two interesting quantities can be derived from the prices,
namely the volatility and the spread.
Yet, computing these quantities requires a bit of care.
In order to explain the problem, let us take an order queue with the best bid at 49,
the best ask at 51, with the next order on the ask side at 151.
The mid price is at 50, with a relative spread of 4\%.
Then, let us assume that the best ask order in the queue is lifted.
After the transaction, the mid price jumps at 100, the relative spread at 102\%,
and the volatility induced from this price change is very large.
These computed values for the spread and volatility are artificial because,
in the example above,
nobody is willing to match the order in the queue at the price of 151.
This generic problem occurs when the queue is ``empty like'' on a given side,
and the best order is executed:
the next order can be arbitrarily far away, creating artificial
large values for the spread and volatility.

In order to avoid these large values, we define the {\em prevailing} prices.
The idea is that the prevailing bid and ask are the best approximations
of the prices at which people are willing to trade on the bid and ask sides.
In the example above, the prevailing ask stays at 51 after the transaction,
as someone did trade at this price. 
If afterward a trader puts a buy order in the queue at a price of 50,
the prevailing bid goes at 50.
A minimal spread is enforced in the algorithm, 
to keep the constraint that the prevailing bid is lower than the prevailing ask.
Continuing our example, with a predefined minimal spread of 1\%,
a new buy order at 51 will put the prevailing bid at 51,
while the prevailing ask is pushed at 51.5.
These ``empty like queue'' situations occur infrequently,
and therefore the details of the algorithm and the value of the parameters
are not very significant, as long as the very large artificial jumps
of a ``naive'' computation with the best bid and ask are avoided.
In the actual computations, the minimum spread as set by the LSE,
and given in Eq.~\ref{eq:minPriceIncrement}, has been used.

All the formula below are defined in terms of the prevailing prices,
but in order to alleviate the notation, it is not reflected in the symbols.
The (prevailing) (logarithmic) spread $s_i$ and (prevailing) (logarithmic)
mid-price $x_i$ are straightforward to define
\begin{eqnarray}
	s_i & = & 10000 \left[\ln(\ask_i) - \ln(\bid_i)\right]  \\
	x_i & = & 0.5 \left[ \ln(\ask_i) + \ln(\bid_i) \right]
\end{eqnarray}
where the spread is measured in BIPS.
The mean spread and logarithmic mid-price are computed with a simple arithmetic average.
The mean price $p$ is computed with a geometric mean
\begin{equation}
   p = \exp( \avg{x} ).
\end{equation}

With high frequency data, the volatility can be measured in several ways.
The annualized volatility is computed, at time $t$, on a time interval $\Delta t$, as follows:
\begin{equation}
  \annVol^2(t) = 100 \sqrt{\frac{1y}{\Delta t}} \sum_{t - \Delta t < t_j \leq t} \left( x_j - x_{j-4} \right)^2
\end{equation}
and is measured in \%. The symbol $1y$ denotes a one year time interval.
From the prevailing mid-prices $x_i$, the price differences are computed 4 ticks apart
in order to lower the very short term volatility due to the individual transactions
(taking price differences 4 ticks apart is arbitrary,
but we do not expect the scaling dependency
to depend on this particular choice).
The factor $\sqrt{1y/\Delta t}$ is used to annualize the volatility (using a random walk hypothesis).
For the actual computation, we have used $\Delta t = $15 minute.
This formula is essentially a generalization to inhomogeneous high frequency
data of the usual formula using daily data.
Then, the daily mean for the annualized volatility is computed using a simple arithmetic average.

Another measure of volatility is the mean tick-by-tick volatility, defined by
\begin{equation}
  \tickVol[p](t) = 10000 \left\{ \frac{1}{n}
  	\sum_{t - \Delta t < t_j \leq t} \left| x_j - x_{j-1} \right|^p \right\}^{1/p}
\end{equation}
and is measured in BIPS.
The sum runs over all the ticks in a time interval $\Delta t$, and $n$ is the number of ticks
included in the sum.
This volatility measures the typical (logarithmic mid-) price changes between two consecutive ticks.
The overnight price differences between closing and opening are not included in the sum.
We have computed $\tickVol[p]$ with $p=1$ and $p=2$.
For all the stocks, both tick volatilities are approximately related with
$\tickVol[2] \simeq 4 \tickVol[1]$.
Therefore, only $\tickVol[2]$ is used for the scaling analysis.

Measures of volatility at very high frequency
are known to be sensitive to microstructure noise,
and are (strongly) biased upward when using trade prices because of the spread
(see e.g. \cite{corsi.2001} for a discussion with foreign exchange data,
and \cite{russell.2003} with stocks data).
By using prevailing bid and ask prices to compute the (prevailing) mid-prices,
we eliminate the random bid-ask bounce responsible for the
upward bias when computing volatility from transaction data.
These mid-prices enter in our estimators for the volatilities,
resulting in an estimate for the annualized volatility which is roughly consistent
with the one computed at lower frequency (say with hourly returns).

\begin{figure}[htb]
  \centering
  \includegraphics[width=0.8\textwidth]{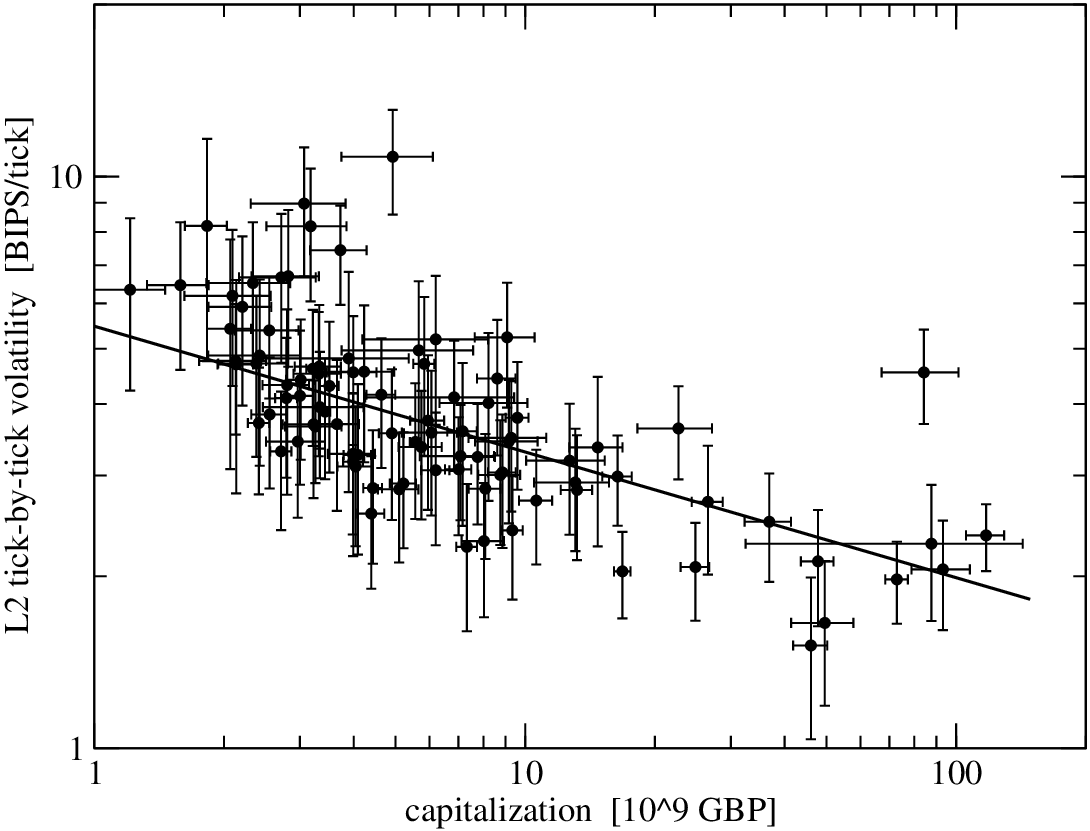}
  \caption{\small \sf The tick-by-tick volatility $\tickVol$ versus the market capitalization $M$.
   The continuous line corresponds to $\tickVol = 5.5 \left( M \right)^{-0.23}$.
  }
  \label{fig:L2tickVolatilityVsCap}
\end{figure}

\begin{figure}[htb]
  \centering
  \includegraphics[width=0.8\textwidth]{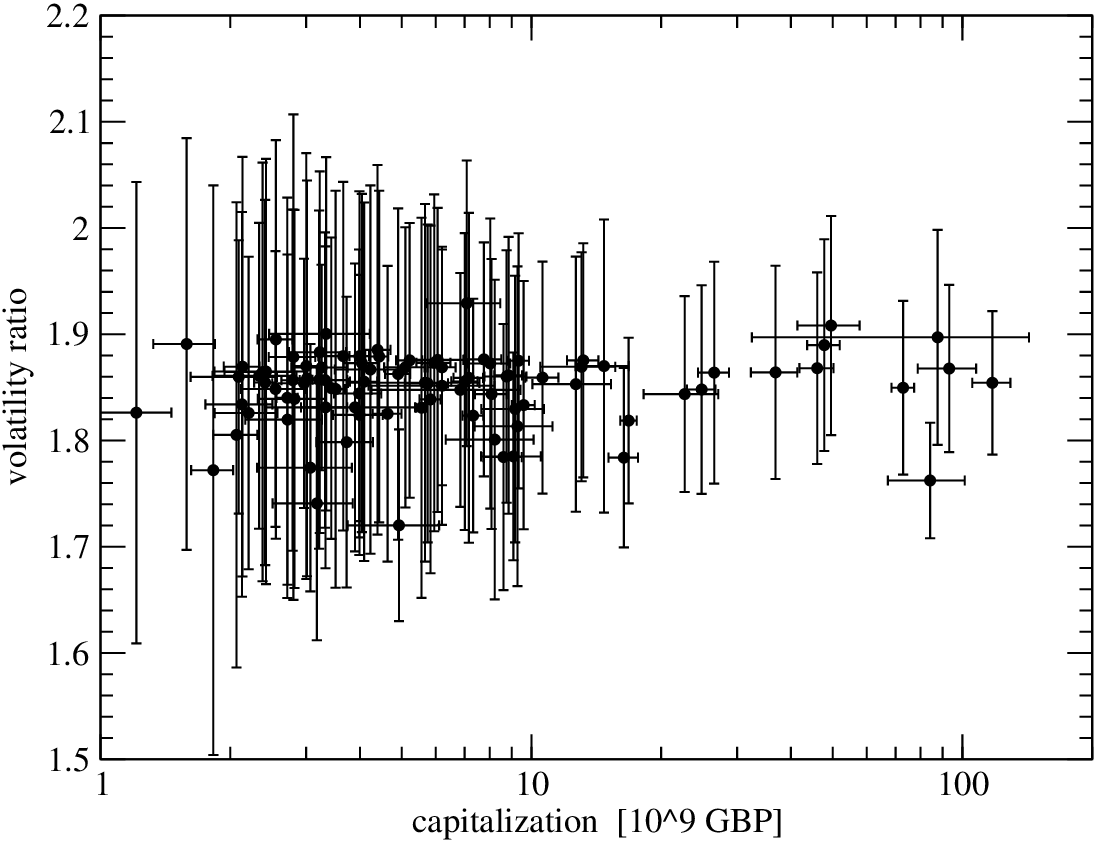}
  \caption{\small \sf The volatility ratio $c$ versus the market capitalization.
  }
  \label{fig:volatilityRatioVsCap}
\end{figure}

Fig.~\ref{fig:L2tickVolatilityVsCap} shows that the tick-by-tick
volatility decreases with an increasing capitalization,
with the exponent of -0.23 $\pm$ 0.02.
The annualized volatility $\annVol$ shows no clear dependency on the market capitalization $M$,
and the least square estimate gives an exponent of 0.04 $\pm$ 0.03, compatible with 0.
The exponents for these two scaling laws can be related from the relation between volatilities
\begin{equation}
   \annVol = c ~\sqrt{dD/dt} ~\tickVol[2] \label{eq:volatilityRelation}
\end{equation}
that can be deduced assuming a random walk for the price.
The constant $c$ depends on the probability distribution for the returns and the various units.
This constant can be extracted from the relation \ref{eq:volatilityRelation},
leading to $c = \annVol/(\sqrt{dN/dt} ~\tickVol[2])$
where the right hand side contains only empirical quantities.
The empirical values for $c$ are reported in Fig.~\ref{fig:volatilityRatioVsCap}.
With the units we are using ($\annVol$ in \%, tick rate in tick/hour, $\tickVol$ in BIPS),
the theoretical value for $c$ is $100/\sqrt{8760} = 1.07$,
whereas a typical value of 1.7 is observed.
Given the above values for the scaling exponent of the tick rate and tick-by-tick volatility,
the relation \ref{eq:volatilityRelation} between volatilities gives a scaling exponent 0.02
for the annualized volatility,
which is compatible with the value computed with the LSQ estimate
for the annualized volatility,
and which is compatible with 0.
In short, we find that
the annualized volatility is independent of the capitalization for the stock included in the FTSE100.
This is contrary to the common belief,
where a decreasing volatility is expected with an increasing capitalization.
The scaling for the annualized volatility was already studied by \cite{plerou.1999} for NYSE data.
They also found that for the companies with the highest capitalizations,
the (annualized) volatility is independent of the market capitalization.
For the smallest capitalization, another regime is observed with a negative exponent
(the smaller the company, the larger the volatility).
With the FTSE data, a larger set of stocks
is probably needed to observe this cross-over.

\begin{figure}[htb]
  \centering
  \includegraphics[width=0.8\textwidth]{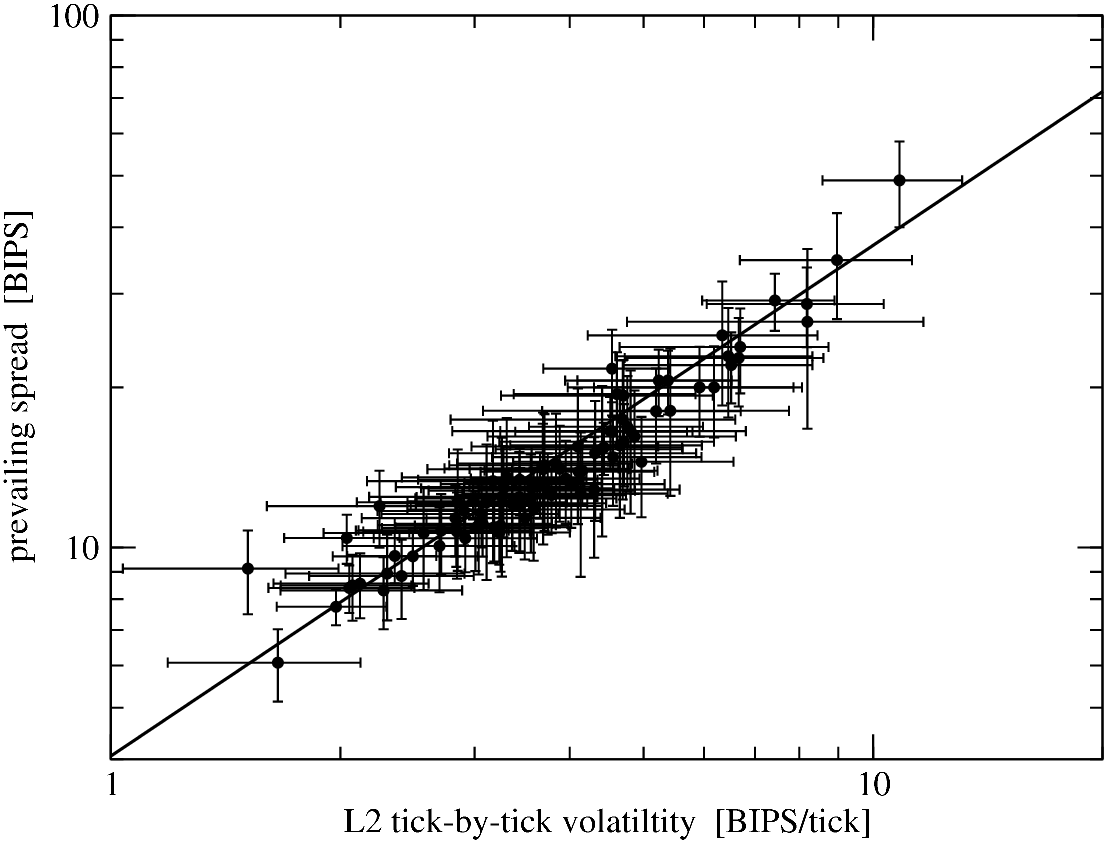}
  \caption{\small \sf The spread versus the tick-by-tick volatility $\tickVol[2]$.
   The continuous line corresponds to $s = 4 ~\tickVol[2]^{0.96}$.
  }
  \label{fig:spreadVsTickVolatility}
\end{figure}

With respect to a scaling analysis, the spread $s$ is explained by two factors.
First, the spread is tighter for lower volatility, as shown in Fig.~\ref{fig:spreadVsTickVolatility}.
The scaling exponent is 0.94 $\pm$ 0.08, compatible with an exponent of 1.
The explanatory power of the tick-by-tick volatility is very good,
but the span for the volatility is of only one decade,
which makes the exponent estimation less robust.
The best correlation with the spread is obtained
with the tick-by-tick volatility $\tickVol[p]$,
whereas the annualized volatility shows a larger
dispersion and an exponent of 1.6 $\pm$ 0.25.
A first result in this direction, but using daily data,
was obtained by \cite{roll.1984} who find a positive correlation
between an indirect measure of the spreads and company sizes.
Recently, \cite{russell.2003} developed a new estimator for the spread
based on transaction prices.
Using S\&P 100 data, in a multiple cross-sectional regression
between spreads and several candidate explanatory factors,
they found a positive regression coefficient with the volatility.
All these results are consistent with microstructure theories
that predict an increasing spread with increasing
volatility (see e.g. \cite{russell.2003} for a discussion of these theories).
The present empirical result indicates that the dependency is
a simple proportionality between spread
and tick-by-tick volatility.

\begin{figure}[htb]
  \centering
  \includegraphics[width=0.8\textwidth]{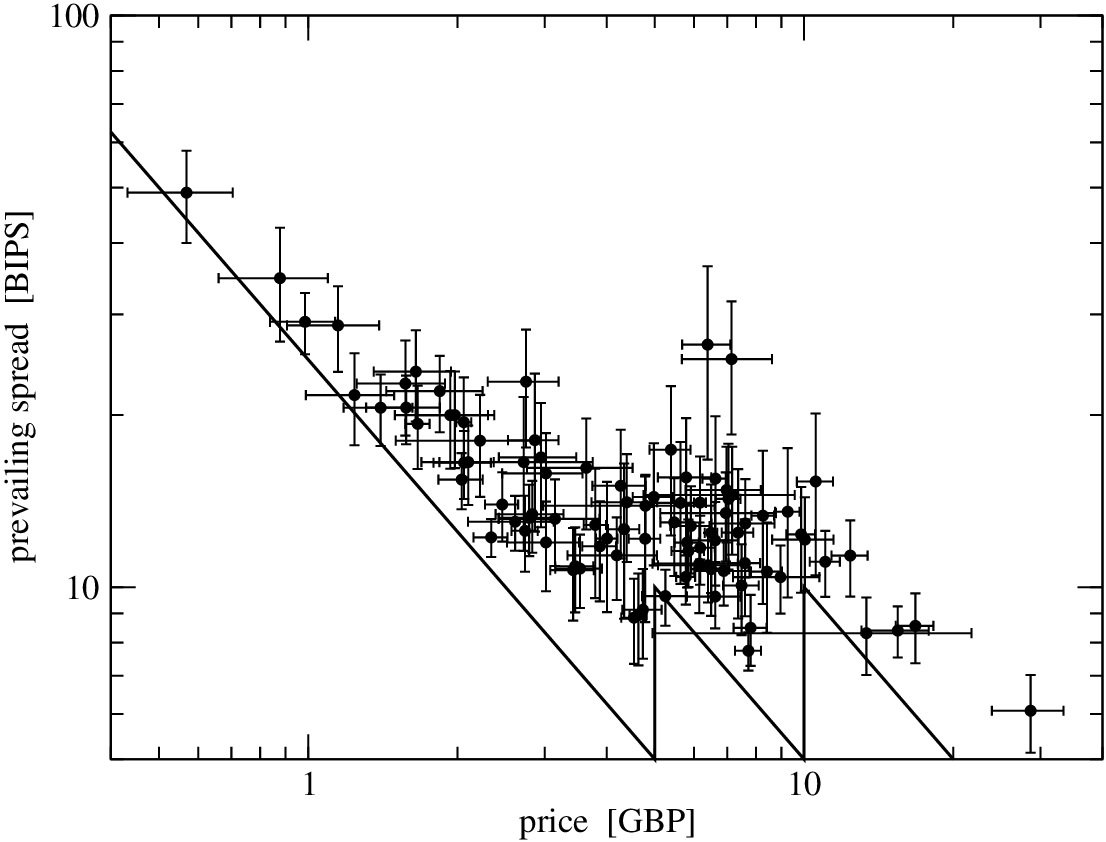}
  \caption{\small \sf The mean spread $s$ versus the mean price $p$.
   The bold jaggy line is the minimal relative price increment, as defined by the London Stock Exchange.
  }
  \label{fig:spreadVsPrice}
\end{figure}

Somewhat unexpectedly, a second factor explaining the mean spread
is the mean price, given on Fig.~\ref{fig:spreadVsPrice}.
Obviously, the minimal spread as set by the LSE rules has a strong
influence on the mean (logarithmic) spread.
Quantitatively, the least square estimate rejects a simple scaling because of the poor goodness of fit.

\section{Discussion and conclusions}
\label{sec:conclusions}

For the quantities measuring the trading activity,
with an increasing mean capitalization $M$, we expect an increase
in the tick rate $dN/dt$, the value per transaction $dV/dD$
and the transaction value per hour $dV/dt$.
At first sight, we may expect all of them to be proportional to
the mean capitalization, but this would lead to a contradiction
due to the scaling relation derived from Eq.~\ref{eq:scalingRelation}.
On a second thought, things are more subtle.
Large pension funds represent a large fraction of the
market (in term of the value they hold).
If their orders dominate the intra-day market,
and if they duplicate simply the index,
we would expect the value per transaction and the transaction
value per hour to be proportional to the market capitalization,
whereas the tick rate and the transaction per tick to be independent of the
market capitalization.
However, traders are known to split large orders to minimize the impact on the market
(but it is in general not said how they quantitatively split the orders to achieve that goal).
This reduces the value per transaction and increases the tick rate,
so that both become non trivial functions of the market capitalization.
Yet, the intra-day market is likely to be dominated
by intra-day speculators and short term portfolio managers.
These market participants use short term criterion to evaluate
profit and risk.
For example, intra-day speculators could prefer
stocks with high liquidity and tighter spread,
in order to minimize the transaction cost.
As a high liquidity corresponds to a high market capitalization,
we can expect that the resulting tick rate is a convex function of the market capitalization.
Then, a full picture of the market must include traders working at different time horizons,
or working with specific stocks or sectors.
However their repartition across stocks and their trading strategies are quantitatively unknown.

\begin{table}\begin{center}
\begin{tabular}{|r|l|l|l|l|l|l|}
\hline
                                      &         & slope    &           & intercept  &            & goodness \\
   ordinate                           &  slope  & std.dev. & intercept & std.dev.   &  $\chi^2$  & of fit   \\
\hline
nb of shares                          &  1.32   &  0.02    &   4.71    &   0.03     &     3193       & 0             \\
prevailing log. spread $s$            & -0.22   &  0.02    &   3.06    &   0.04     &      212       & $7~10^{-12}$ \\
\hline
tick rate             $dN/dt$         &  0.39   &  0.03    &    5.17   &   0.07     &        40.0     &  1          \\
trsct value per tick  $dV/dN$         &  0.48   &  0.06    &   -6.53   &   0.14     &        15.6     &  1          \\
trsct value per trsct $dV/dD$         &  0.44   &  0.03    &   -4.72   &   0.07     &        28.0     &  1          \\
nb trsct per tick     $dD/dN$         &  0.05   &  0.05    &   -1.95   &   0.09     &        11.3     &  1          \\
trsct value per hour  $dV/dt$         &  0.90   &  0.055   &   -1.62   &   0.12     &        32.3     &  1          \\
\hline
nb order in queue    $n_\text{queue}$ &  0.32   &  0.03    &  -2.08    &   0.06     &         67.6    &  0.96     \\
order value in queue $v_\text{queue}$ &  0.76   &  0.04    &  -6.11    &   0.09     &        60.1     &  0.99     \\
\hline
annual volatility        $\annVol$    & -0.03   &  0.03    &   4.91    &   0.07     &        73.5     & 0.90      \\
tick volatility (p = 1) $\tickVol$    & -0.21   &  0.03    &   0.31    &   0.06     &         92.2    &  0.42     \\
tick volatility (p = 2) $\tickVol$    & -0.23   &  0.02    &   1.75    &   0.06     &        118      & 0.02      \\
volatility ratio           $c$                & -0.00   &  0.006   &   0.61    &   0.02     &          9.61   &    1      \\
\hline
\end{tabular}
\caption{\label{table:meanCap} Least square estimates for the quantities given in the first column,
    with respect to the  capitalization $M$.
    Notice the the first 2 relations are rejected based on their poor goodness of fit,
    the one involving the tick-by-tick volatility with $p=2$ is marginal,
    whereas all the others are unambiguously accepted.
 }
 \end{center}
\end{table}
\begin{table}\begin{center}
\begin{tabular}{|r|r|l|l|l|l|l|l|}
\hline
                               &                   &                 & slope        &                & intercept   &                  & goodness \\
      abscisse          &   ordinate   &  slope       & std.dev.    & intercept & std.dev.     &   $\chi^2$  & of fit   \\
\hline
tick volatility (p=2)  & prevailing log. spread & 0.94  &   0.08  &   1.40   &   0.11    &   11.9 &    1    \\
nb order in queue      & order value in queue   &  2.14  &   0.18 &   -1.50   &   0.28   &  13.1   &    1    \\
order value in queue   & trsct value per trsct  &  0.57  &   0.05 &   -1.20   &   0.22   &  16.4   &   1    \\
\hline
\end{tabular}
\caption{\label{table:otherLSQ} Least square estimates for the quantities given in the second column,
    with respect to the one given in the first column.
 }
\end{center}
\end{table}
The outcome of the above scaling analysis
is that all these factors conspire so that clear
scaling relations are observed.
For the scaling relations discussed in this paper,
the full results of the LSQ estimates are given in
table \ref{table:meanCap} for the dependencies with
respect to the mean capitalization, and in table \ref{table:otherLSQ}
for the other dependencies.
First, a simple proportionality between
the transaction value per hour $dV/dt$ and the capitalization could be expected,
and the computed exponent is 0.90 $\pm$ 0.055.
Similarly, the value in the order queue could be expected
to be proportional to the market capitalization,
but an exponent 0.76 $\pm$ 0.04 is measured.
Clearly, similar analysis using data from other stock exchanges should be done.
Second, the tick rate $dN/dt$, the transaction value per tick $dV/dD$,
and the number of transactions per tick $dD/dN$ have
scaling exponents 0.37 $\pm$ 0.03, 0.44 $\pm$ 0.03, and 0.05 $\pm$ 0.05 respectively.
These three scaling relations combined together imply a scaling exponent 0.86 for
the transaction value per hour $dV/dt$, in line with the estimated exponent 0.90.
Otherwise, the scaling analysis for the order queue is  consistent
with the behavior of the transactions.

Turning to the prices and their dynamics, we observe that the mean logarithmic spread
has a large correlation with the tick-by-tick volatility, with an exponent 0.94.
Intuitively, we expect that the spread grows with the volatility,
and empirically a near proportionality is observed.

The annualized volatility $\annVol$ is independent of the market capitalization,
in agreement with the relation  \ref{eq:volatilityRelation}
and the empirical scaling relation for the
tick rate and tick-by-tick volatility.
On the other hand, the tick-by-tick volatility decreases with an increasing capitalization,
with the exponent -0.23.
This value for the exponent can be obtained from the following heuristic argument.
The values in the order queue at the best bid and ask scale
with the market capitalization $v_\text{queue} \simeq M^{0.76}$
and the value per transaction scales as
$dV/dD \simeq M^{0.44}$.
Assuming a random walk behavior for the transactions and the new orders,
the number of transactions $n$ needed to completely empty
the queue at the best bid or ask positions is given by $n^2 dV/dD = v_\text{queue}$.
Using the observed values for the exponents, we obtain $n \simeq M^{0.16}$.
For n-1 ticks, the best positions are not empty and the mid-price does not change,
and then the mid-price changes by one price increment every $n$-th tick.
Therefore, the mean tick-by-tick volatility is proportional to $1/n$,
and $\tickVol \simeq M^{-0.16}$, roughly in agreement with the empirical exponent -0.23.
Essentially, this argument links the scaling for the tick-by-tick volatility to the scaling
for the value per transaction and value in the queue,
using a random walk relation for the number of orders needed to change the best price.
This is the most interesting argument as it relates the exponents in the 3 groups,
relating the dynamic of the incoming orders, the orders in the queue and the price volatility.

\bibliographystyle{apalike}
\bibliography{FTSEscalingWithSize}

\end{document}